\documentclass[12pt]{article}
\usepackage{epsfig,psfrag}
\usepackage{citesort}
\date{\today}
 \normalsize

\newcommand{\bmat}{\left(\begin{array}}
\newcommand{\emat}{\end{array}\right)}
\newcommand{\be}{\begin{equation}}
\newcommand{\ee}{\end{equation}}
\newcommand{\bea}{\begin{eqnarray}}
\newcommand{\eea}{\end{eqnarray}}
\def\ie{{\it i.e.}}
\def\lsim{\raise0.3ex\hbox{$\;<$\kern-0.75em\raise-1.1ex\hbox{$\sim\;$}}}
\def\gsim{\raise0.3ex\hbox{$\;>$\kern-0.75em\raise-1.1ex\hbox{$\sim\;$}}}

\begin{document}
\renewcommand{\thefootnote}{\fnsymbol{footnote}}
\begin{titlepage}
\rightline{\today} \vspace{.5cm}
{\large
\begin{center}
{\large \bf  Supersymmetric contributions to $B\to D K$ and the
determination of angle $\gamma$}
\end{center}}
\vspace{.1cm}
\begin{center}
Shaaban Khalil\\

\vspace{.3cm} \emph{Ain Shams University, Faculty of Science,
Cairo 11566,
Egypt.}\\
\emph{German University in Cairo, New
Cairo city, El Tagamoa El Khames, Egypt.}\\
\end{center}
\vspace{.5cm}
\begin{minipage}[h]{14.0cm}
\begin{center}
\small{\bf Abstract}\\[3mm]
\end{center}
We analyze supersymmetric contributions to the branching ratios
and CP asymmetries of $B^- \to D^0 K^-$ and $B^- \to \bar{D}^0
K^-$ processes. We investigate the possibility that supersymmetric
CP violating phases can affect our determination for the angle
$\gamma$ in the unitary triangle of Cabibbo-Kobayashi-Maskaw
mixing matrix. We calculate the gluino and chargino contributions
to $b\to u(\bar{c}s)$ and $b\to c(\bar{u}s)$ transitions in a
model independent way by using the mass insertion approximation
method. We also revise the $D^0 - \bar{D}^0$ mixing constraints on
the mass insertions between the first and second generations of
the up sector. We emphasize that in case of negligible $D^0 -
\bar{D}^0$ mixing, one should consider simultaneous contributions
from more than one mass insertion in order to be able to obtain
the CP asymmetries of these processes within their $1\sigma$
experimental range. However, with a large $D^0-\bar{D}^0$ mixing,
one finds a significant deviation between the two asymmetries and
it becomes natural to have them of order the central values of
their experimental measurements.

\end{minipage}
\end{titlepage}

\section{{\large \bf Introduction}}
Recently, the BaBar collaborations have measured the charge CP
asymmetries $A_{CP_{\pm}}$ and the branching ratios $R_{CP_{\pm}}$
of the $B^-\to D^0 K^-$ and $B^- \to \bar{D}^0 K^-$ decays
\cite{Aubert:2005rw}. The following results have been reported:
\bea \!\!\!A_{CP_+}\!&\!=\!&\!0.35 \pm 0.13(\rm{stat}) \pm 0.04
(\rm{syst}),~~~~A_{CP_-} = - 0.06 \pm 0.13(\rm{stat}) \pm 0.04
(\rm{syst}),\label{Exp1}\\
\!\!\!R_{CP_+}\!&\!=\!&\!0.90 \pm 0.12(\rm{stat}) \pm 0.04
(\rm{syst}),~~~~R_{CP_-} = 0.86 \pm 0.10(\rm{stat}) \pm 0.05
(\rm{syst}).\label{Exp2} \eea These results, with all other
$B$-factories measurements, provide a stringent test of the Standard
Model (SM) picture of flavor structure and CP violation and open the
possibility of probing virtual effect from new physics at low
energy.

In the SM, CP violation arises from complex Yukawa couplings which
lead to the angles $\alpha,\beta$ and $\gamma$ in the unitary
triangle of Cabibbo-Kobayashi-Maskawa (CKM) quark mixing matrix. The
angle $\beta = \mathrm{arg}\left(-\frac{V_{cd} V^*_{cb}}{V_{td}
V^*_{tb}}\right)$ has been determined by the CP asymmetry in $B^0
\to J/\psi K_S$ process which is dominated by tree level
contribution. Concerning the angle
$\gamma=\rm{arg}\left(-\frac{V_{ud}V_{ub}^*}{V_{cd}V_{cb}^*}\right)$,
it is believed that a theoretically clean measurement of this angle
can be obtained from exploiting the interference between $B^-\to D^0
K^-$ and $B^- \to \bar{D}^0 K^-$ when $D^0$ and $\bar{D}^0$ mesons
decay to the same CP eigenstate \cite{Gronau:1990ra}.

At the quark level, the $B^- \to \bar{D}^0  K^-$ and $B^- \to D^0
K^-$ decays are based on $b \to u (\bar{c}s)$ and $b \to c
(\bar{u}s)$ transitions respectively. Therefore, their SM
contributions at tree leve are suppressed by the CKM factors
$V_{cs} V^*_{ub}$ and $V_{us}^* V_{cb}$ which are of order
$10^{-3}$. This gives the hope that it may be possible for a new
physics beyond the SM, like supersymmetry, which contributes to
these decays at one loop level to manifest itself and compete the
SM. In this paper we aim to investigate this possibility and
check, in a model independent way, whether supersymmetry can
significantly modify the CP asymmetries in $B^- \to D K^-$
processes and hence affects the determination of the angle
$\gamma$. Therefore, we perform a systematic analysis of the SUSY
contributions to $B\to D K$ processes. We compute SUSY
contributions to $b \to u (\bar{c}s)$ and $b \to c (\bar{u}s)$
transitions through the gluino and chargino exchange, using the
mass insertion approximation method. This approximation is quite
useful tool for studying the SUSY contributions to the flavor
processes in a model independent way. We show that the gluino box
diagrams give the dominant SUSY contribution while the chargino
exchanges lead to subdominant contributions.

It turns out that the $D^0 - \bar{D}^0$ mixing may limit the gluino
contribution to $B^-\to D K^-$ due to the stringent constraints on
the mass insertions between the first and second generations in the
up sector, $(\delta^u_{AB})_{12}$. Thus in our analysis, we revise
the $D^0 - \bar{D}^0$ mixing constraints \cite{Chang:2001ah} and
take them into account. We find that with a single mass insertion,
the SUSY contribution to $B^-\to D K^-$ decay will be much smaller
than the SM result. Nevertheless, with simultaneous contributions
from more than one mass insertion, the SUSY effect can be enhanced
and the results of the CP asymmetries become within $1\sigma$
experimental range, while the $D^0 - \bar{D}^0$ mixing constraints
are satisfied.

The paper is organized as follows. In section 2 we study the CP
asymmetries and the branching ratios of $B^- \to D K^-$ in the SM.
We show that in the SM the branching ratios $R_{CP_{\pm}}$ are
within the experimental range. While the CP asymmetry $A_{CP_{+}}$
is below its $1\sigma$ experimental lower bound and the value of
$A_{CP_{-}}$ is typically negative. In section 3 we compute the
gluino and chargino contributions to $b\to u(\bar{c}s)$ and $b\to
c(\bar{u}s)$ transitions in terms of the mass insertions. Section
4 is devoted for analyzing the SUSY contribution to
$D^0-\bar{D}^0$ mixing and revise the possible constrain on the
mass insertions $(\delta^u_{AB})_{12}$. The analysis of SUSY
contribution to CP asymmetries $A_{CP_{\pm}}$ and branching ratios
$R_{CP_{\pm}}$ is given in section 5. We show that in case of
negligible $D^0 - \bar{D}^0$ mixing, one should consider
simultaneous contributions from more than one mass insertion in
order to obtain $A_{CP_{\pm}}$ within their $1\sigma$ experimental
range. Nevertheless, the usual relation: $A_+ \simeq - A_-$ which
is valid in the SM remains hold. With a large $D^0-\bar{D}^0$
mixing, one finds a significant deviation between $A_+$ and $A_-$
and it becomes natural to obtain $A_{CP_{\pm}}$ of order the
central values of their experimental measurements. Finally, we
give our conclusions in section 6.

%
\section{{\large \bf $B^- \to D K^-$ in the Standard Model}}
\begin{figure}[h]
\epsfig{file=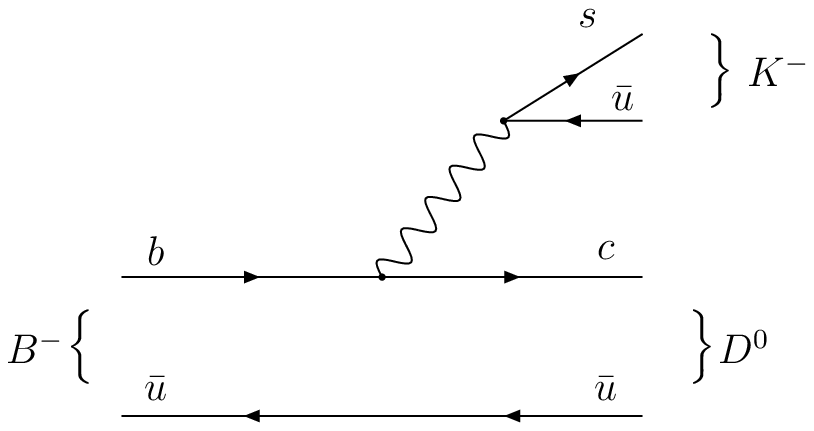, width=7cm, height=3cm, angle=0}
\hspace{0.5cm}\epsfig{file=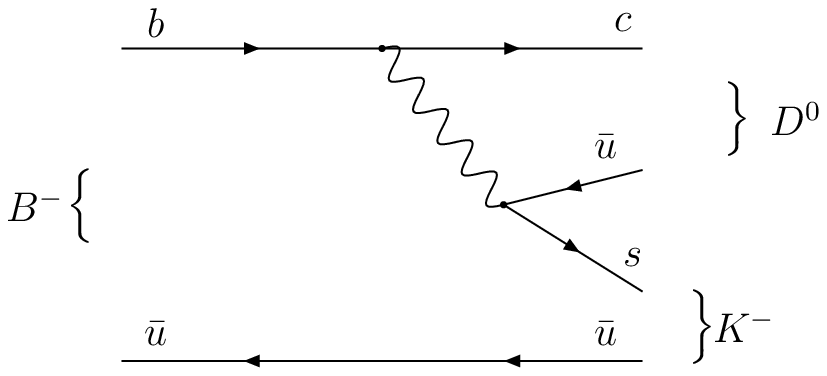, width=7cm, height=3cm,
angle=0}\\
\vskip 0.5cm \epsfig{file=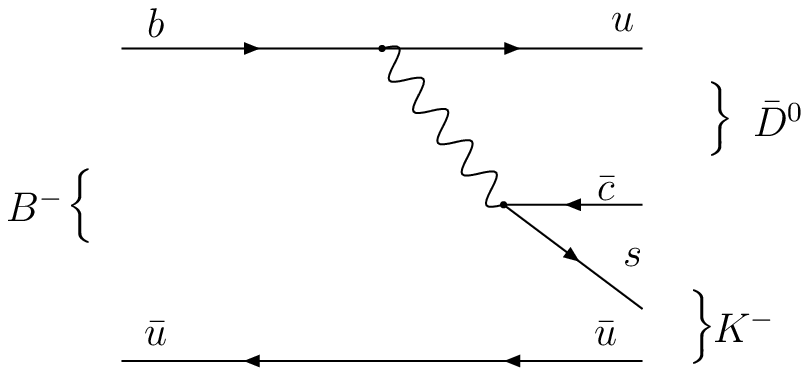, width=7cm, height=3cm,
angle=0}\hspace{0.5cm}\epsfig{file=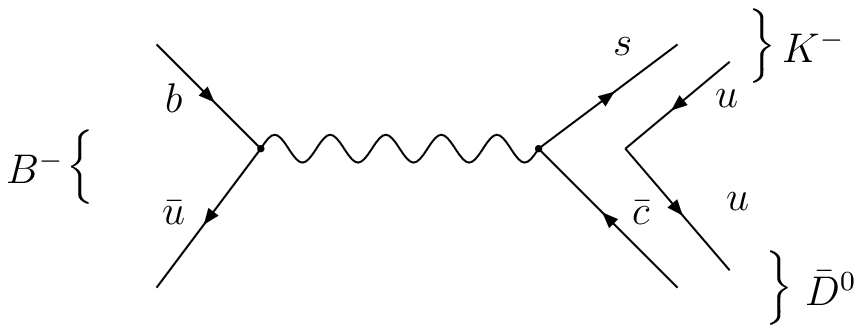, width=7cm,
height=3cm, angle=0} \caption{SM contributions to $B^-\to D K^-$:
color-favored tree (left up), color-suppressed tree (right-up and
left-down) and annihilation (right-down)} \label{SMfig}
\end{figure}

In this section we analyze the CP violation in $B^- \to D K^-$
decays within the SM. The possible quark level topologies of $B^-\to
D K^-$ that contribute to the amplitude $A(B^-\to D^0 K^-)$ and
$A(B^- \to \bar{D}^0 K^-)$ in the SM can be classified to the
following three categories, as shown in Fig.(\ref{SMfig}):
color-favored tree (T), color-suppressed tree (C) and annihilation
(A).
These processes  are given in terms of the CKM factors $\lambda_c=
V_{cb} V_{us}^*$, $\lambda_u= V_{ub} V^*_{cs}$. The decay $B^- \to
D^0 K^-$ receives contributions from $T$ and $C$ with factor
$\lambda_c$, while $B^- \to \bar{D}^0 K^-$ get contributions from
$C$ and $A$ in terms of $\lambda_u$. Since the contributions from
the annihilation process to the matrix elements are quite suppressed
at the leading order correction \cite{Xing:1995cd}, it is quite
reasonable to assume that $A=0$. In our analysis we will adopt this
approximation and therefore the general parametrization of the SM
amplitudes of $B^- \to D K^-$ decays can be given by \bea
A^{SM}(B^-\to D^0 K^-) &=& \vert A_1 \vert e^{i\delta_1} \equiv
\bar{T} +
\bar{C},\label{D0}\\
A^{SM}(B^-\to \bar{D}^0 K^-) &=& \vert A_2 \vert e^{i\delta_2}
e^{i \gamma} \equiv C  ,\label{D0bar} \eea where $\delta_i$,
$i=1,2$ are the strong (CP-conserving) phases. $\bar{T}$ and
$\bar{C}$ refer to the color allowed and color suppressed tree
amplitudes involving $b\to c(\bar{u}s)$ while $C$ is related to
the process $b\to u (\bar{c}s)$. In terms of the two
CP-eigenstates of the neutral $D$ meson system, $D^0_{CP_{\pm}}=
(D^0 \pm \bar{D}^0)/\sqrt{2}$, one considers the ratios
$R_{CP_{\pm}}$ of charged averaged partial rates and the charge
asymmetries $A_{CP_{\pm}}$: \bea R_{CP_{\pm}}&=&
\frac{2\left[\Gamma(B^- \to D^0_{CP_{\pm}} K^-) + \Gamma(B^+ \to
D^0_{CP_{\pm}} K^+)\right]}{\Gamma(B^- \to D^0
K^-)+\Gamma(B^+ \to \bar{D}^0 K^+)}\\
A_{CP_{\pm}}&=& \frac{\Gamma(B^- \to D^0_{CP_{\pm}} K^-) -
\Gamma(B^+ \to D^0_{CP_{\pm}} K^+)}{\Gamma(B^- \to D^0_{CP_{\pm}}
K^-)+\Gamma(B^+ \to D^0_{CP_{\pm}} K^+)}. \eea We define the ratio
of the SM amplitudes of $B^-\to \bar{D}^0 K^-$ and $B^-\to D^0 K^-$
as \be r_B e^{i\delta_B} e^{i\gamma} = \frac{A^{SM}(B^-\to \bar{D}^0
K^-)}{A^{SM}(B^-\to D^0 K^-)}. \label{ratio}\ee According to
Eqs.(\ref{D0},\ref{D0bar}), $r_B=\vert A_2/A_1\vert$ and $\delta_{B}
= \delta_2 -\delta_1$. Using this parametrization, one finds that
$R_{\pm}\equiv R_{CP_{\pm}}$ is given by \be R_{\pm} = 1 + r_B^2 \pm
2 r_B \cos\delta_B \cos \gamma, \label{RCP}\ee and $A_{\pm}\equiv
A_{CP_{\pm}}$ takes the form \be A_{\pm} = \frac{\pm 2 r_B
\sin\delta_B \sin \gamma}{1 + r_B^2 \pm 2 r_B \cos\delta_B \cos
\gamma} \equiv \frac{\pm 2 r_B \sin\delta_B \sin
\gamma}{R_{CP_{\pm}}}\label{ACP}. \ee From Eq.(\ref{RCP}) one gets
\be \cos \gamma = \frac{R_+ - R_-}{4 r_B \cos \delta_B}.\ee Thus, by
using the expressions for the CP asymmetries $A_{\pm}$ in
Eq.(\ref{ACP}), one can factorize the dependence on the strong phase
and gets the following expression for the angle $\gamma$ in terms of
$R_{\pm}$, $A_{\pm}$ and $r_B$ only: \be \sin \gamma = \frac{2
\cos\gamma \left(A_+ - A_-\right)}{\sqrt{16 r_B^2 \cos^2\gamma- (R_+
-R_-)^2}} \frac{R_+R_-}{R_+ + R_-}. \ee From this expression, one
can easily see that the central experimental values of $R_{\pm}$ and
$A_{\pm}$ with $r_B\simeq 0.1$ implies that the angle $\gamma$ is of
order $\gamma \simeq 71^{\circ}$. It is worth mentioning that within
the SM, the effect of the $D^0-\bar{D}^0$ mixing is very small on
extracting the angle $\gamma$ using the $B^- \to D K^-$ decays. As
emphasized in Ref.\cite{Grossman:2005rp}, neglecting this mixing
implies an error in determining $\gamma$ of order $0.1-1^{\circ}$.

In order to analyze the SM predictions for the $A_{\pm}$ and
$R_{\pm}$ and compare them with the experimental results reported in
Eqs.(\ref{Exp1},\ref{Exp2}), let us consider the SM contributions to
the $b\to u(\bar{c}s)$ and $b\to c(\bar{u}s)$ transitions. As shown
in Fig. \ref{SMfig}, within the SM the $B^-\to D K^-$ are pure
`tree' decays. The effective Hamiltonian of this transition is given
by \be H_{\rm{eff}}^{b\to u(\bar{c}s)} = \frac{G_F}{\sqrt{2}} V_{ub}
V^*_{cs} \left[ C_1(\mu) Q_1^{u} + C_2(\mu) Q_2^{u}\right],
\label{SMHeff}\ee where $C_i$ and $Q^u_i$ are the Wilson
coefficients and operators of this transition renormalized at the
scale $\mu$ with \bea Q_1^{u} = \left(\bar{u}^{\alpha} \gamma_{\mu}
L b^{\alpha}\right)\left(\bar{s}^{\beta} \gamma^{\mu} L
c^{\beta}\right),~~~~~~~~~~~~~~~ Q_2^{u} = \left(\bar{u}^{\alpha}
\gamma_{\mu} L b^{\beta}\right)\left(\bar{s}^{\beta}\gamma^{\mu} L
c^{\alpha}\right),
 \eea
where $L=(1-\gamma_5)$. The effective Hamiltonian for the $b\to
c(\bar{u}s)$ transition can
 be obtained from the effective Hamiltonian in Eq.(\ref{SMHeff}) by exchanging $u \leftrightarrow c$.
The SM results for the corresponding Wilson coefficients are: \be
C_1(m_W)=1-\frac{11}{6} \frac{\alpha_s}{4 \pi},
~~~~~~~~~~~~~~~~~~~~~~~~~~~ C_2 (m_W) =\frac{14 \alpha_s}{16 \pi}.
\ee However, due to the QCD renormalization to the scale $\mu
\simeq m_b$, $C_1$ and $C_2$ get mixed, as will be discussed in
more details in the next section, and one finds \be C_1(\mu) =
1.07, ~~~~~~~~~~~~~~~~~~~~~~ C_2(\mu) =-0.17. \ee

\begin{figure}[h]
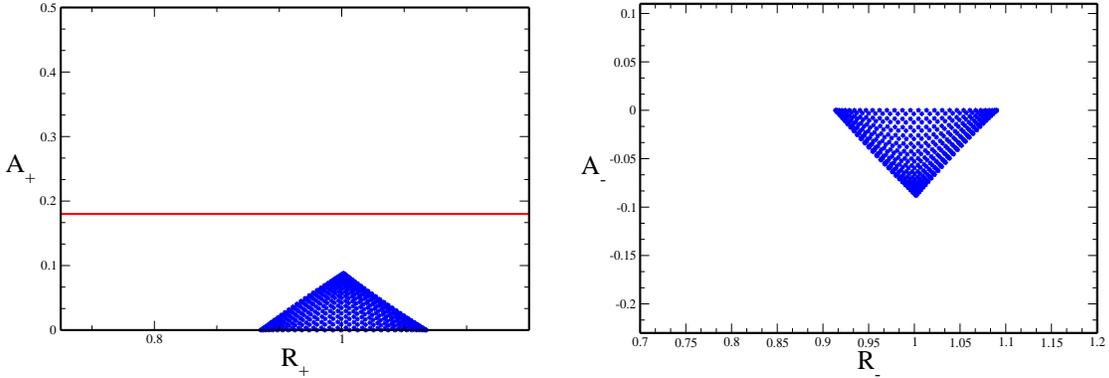

\epsfig{file=RA1.eps, width=7cm, height=5cm, angle=0}
\hspace{0.5cm}\epsfig{file=RA2.eps, width=7cm, height=5cm,
angle=0} \caption{$R_+$ versus $A_+$ and $R_-$ versus $A_-$ within
the Standard Model. The horizontal line in the left figure
represents the lower bound of $A_+$ at $1\sigma$ experimental
range.} \label{SMresult}
\end{figure}

To evaluate the SM results to the decay amplitude of $B^- \to D
K^-$, we have to determine the matrix elements for the operators
$Q_{1,2}^{u,c}$. A detailed analysis for the matrix elements will
be given in the next section. Here, we just give the matrix
elements for these four operators in naive factorization: \bea
\langle \bar{D}^0 K^- \vert Q_1^u \vert B^- \rangle &=& -
\frac{X}{3},~~~~~ ~~~~~~~~ ~~~~~~~~\langle \bar{D}^0
K^- \vert Q_2^u \vert B^- \rangle = - X,\\
\langle D^0 K^- \vert Q_1^c \vert B^- \rangle &=&- \frac{1}{3}X
-Y,~~~~~ ~~~~~~~~~\langle D^0 K^- \vert Q_2^c \vert B^- \rangle
=-X - \frac{1}{3} Y , \eea where \be X = i F_0^{B\to K}(m_D^2) f_D
(m_B^2 - m_K^2),~~~~~~ Y = i F_0^{B\to D}(m_K^2) f_K (m_B^2 -
m_D^2). \label{XY}\ee There are two comments in order: i) The
naive factorization can not determine the strong phases,
therefore, in our analysis we consider these phases as free
parameters. ii) As mentioned above, the factorized matrix element
$\langle \bar{D}^0 K^- \vert (\bar{s} \gamma^{\mu} L c) \vert
0\rangle \langle 0 \vert \bar{u} \gamma_{\mu} L b \vert B^-
\rangle$ corresponding to an annihilation process is suppressed as
showed in Ref.\cite{Xing:1995cd}, and can be neglected. Therefore,
\be A^{SM}(B^-\to \bar{D}^0 K^-) = - \frac{G_F}{\sqrt{2}}
V_{ub}V_{cs}^* X \left(\frac{C_1}{3} + C_2\right),\ee and \be
A^{SM}(B^-\to D^0 K^-) = - \frac{G_F}{\sqrt{2}} V_{cb}V_{us}^*
\left[X \left(\frac{C_1}{3}+ C_2\right) + Y \left(C_1+
\frac{C_2}{3}\right)\right].\ee Fixing the hadronic parameters as
follows: $f_D=0.2$, $f_K=0.16$, $F_0^{B\to D}=0.34$, $F_0^{B\to
K}=0.62$, and the meson masses as: $m_K=0.49$, $m_D=1.86$, and
$m_B=5.278$ GeV. One finds \be r_B \simeq 0.05 \label{rB}\ee Note
that it is customary assumed that with a large uncertainty, the SM
prediction for $r_B$ may be much larger than the above value (can
be ${\mathcal O}(0.1)$, see Ref.\cite{Gronau:1998vg}). Here we
will use the value that we obtained in Eq.(\ref{rB}) as a typical
value for the SM contribution. In order to have a general picture
of the SM predictions for the CP asymmetries $A_{\pm}$ and the
branching ratios $R_{\pm}$, we plot in Fig.\ref{SMresult}
$R_{\pm}$ versus $A_{\pm}$. Here, we vary the parameter $\delta_B$
in the range $[0,\pi]$ and the angle $\gamma$
is also considered between $0$ and $\pi$.\\

As can be seen from the results in Fig.\ref{SMresult}, the SM
predictions for the branching ratios $R_{\pm}$ are within the
$1\sigma$ experimental range. However, the results for the CP
asymmetry $A_+$ are below its experimental lower bound. Also the SM
leads to a negative CP asymmetry $A_-$ which is still consistent
with its experimental results in Eq(\ref{Exp1}), due to the large
uncertainties in these measurement. Therefore, more precise
measurements would be very important in analyzing the SM predictions
for $R_{\pm}$ and $A_{\pm}$ and, hence, in determining the value of
the angle $\gamma$.
%
%
\section{{\large \bf SUSY contributions to $b\to u(\bar{c}s)$ and $b\to c(\bar{u}s)$}}

The crucial point to note from the previous section, is that the SM
contributions to the amplitudes of $b\to u(\bar{c} s)$ and $b\to
c(\bar{u} s)$ transitions are suppressed by the CKM factors
$V_{ub}\simeq {\cal O}(10^{-3})$ and $V^*_{us}V_{cb}\simeq {\cal
O}(10^{-3})$ respectively. Therefore, it may be possible to have a
comparable effect from new physics at one loop level which can
compete with the SM tree level contribution. In this section we
study the supersymmetric contributions to the $b\to u(\bar{c}s)$ and
$b\to c(\bar{c}s)$ transitions. In this case, the effective
Hamiltonian $H_{\rm eff}^{\Delta C=1}$ for the $b \to u (\bar{c}s)$
can be expressed as \be H_{\rm eff}^{\Delta C=1} = \sum_{i=1}^{10}
\left( C^u_i(\mu)~ Q^u_i(\mu)~ +~ \tilde{C}^u_i(\mu) ~
\tilde{Q}^u_i(\mu) \right),\label{effH} \ee where $C^u_i$ are the
Wilson coefficients and $Q^u_i$ are the relevant local operators at
low energy scale $\mu \simeq m_b$. The operators $Q_i^u$ are given
by \bea Q^u_1 &=& \left(\bar{u}^{\alpha} \gamma_{\mu} L
b^{\alpha}\right)\left(\bar{s}^{\beta} \gamma^{\mu} L
c^{\beta}\right),~~~~~~~~~~~~~  Q^u_2 = \left(\bar{u}^{\alpha}
\gamma_{\mu} L b^{\beta}\right)\left(\bar{s}^{\beta} \gamma^{\mu}
L c^{\alpha}\right),\nonumber\\
Q^u_3 &=& \left(\bar{u}^{\alpha}  \gamma_{\mu}L
b^{\alpha}\right)\left(\bar{s}^{\beta} \gamma^{\mu} R
c^{\beta}\right),~~~~~~~~~ ~~~~Q^u_4 = \left(\bar{u}^{\alpha}
\gamma_{\mu} L b^{\beta}\right)\left(\bar{s}^{\beta} \gamma^{\mu}
R c^{\alpha}\right),\nonumber\\
Q^u_5 &=& \left(\bar{u}^{\alpha} L
b^{\alpha}\right)\left(\bar{s}^{\beta} L
c^{\beta}\right),~~~~~~~~~~~~~~~~~~~ Q^u_6 =
\left(\bar{u}^{\alpha} L b^{\beta}\right)\left(\bar{s}^{\beta}L
c^{\alpha}\right),\\
Q^u_7 &=& \left(\bar{u}^{\alpha} L
b^{\alpha}\right)\left(\bar{s}^{\beta} R
c^{\beta}\right),~~~~~~~~~~~~~~~~~~~ Q^u_8 =
\left(\bar{u}^{\alpha} L b^{\beta}\right)\left(\bar{s}^{\beta}R
c^{\alpha}\right),\nonumber\\
Q^u_9 &=& \left(\bar{u}^{\alpha} \sigma_{\mu\nu} L
b^{\alpha}\right)\left(\bar{s}^{\beta} \sigma^{\mu\nu} L
c^{\beta}\right),~~~~~~~~~~ Q^u_{10} = \left(\bar{u}^{\alpha}
\sigma_{\mu\nu} L b^{\beta}\right)\left(\bar{s}^{\beta}
\sigma^{\mu\nu} L c^{\alpha}\right),\nonumber
 \eea where $\alpha$ and $\beta$ refer to the color indices. $L,R$ are given by $(1\mp \gamma_5)$ respectively and
 $\sigma^{\mu\nu}=\frac{i}{2}[\gamma^{\mu},\gamma^{\nu}]$. The operators $\tilde{Q}_i^u$ are obtained from $Q^u_i$
 by the chirality exchange $L \leftrightarrow R$. In the SM, the coefficients $\tilde{C}_i^u$ are
 identically vanish, while in SUSY models, they receive contributions from both gluino and chargino
 exchanges. The corresponding operators for $b \to c (\bar{u}s)$ can be
obtained from the above expression by exchanging $u
\leftrightarrow c$.

\begin{figure}[h]
\epsfig{file=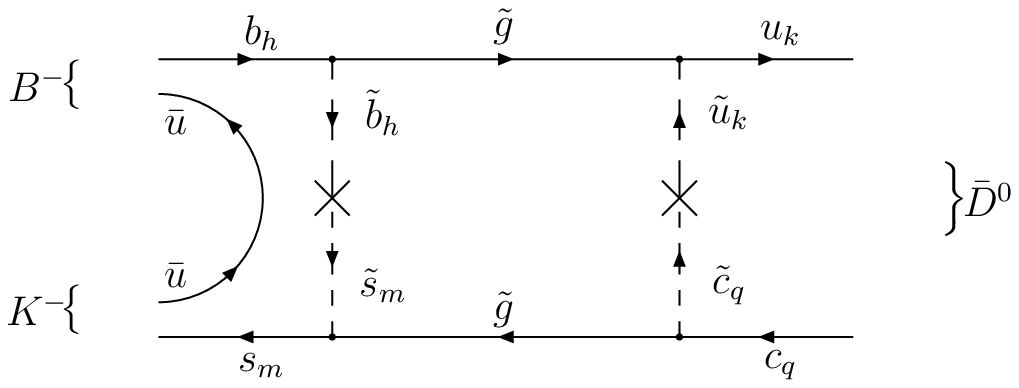, width=8cm, height=3.5cm, angle=0}
\hspace{0.5cm}\epsfig{file=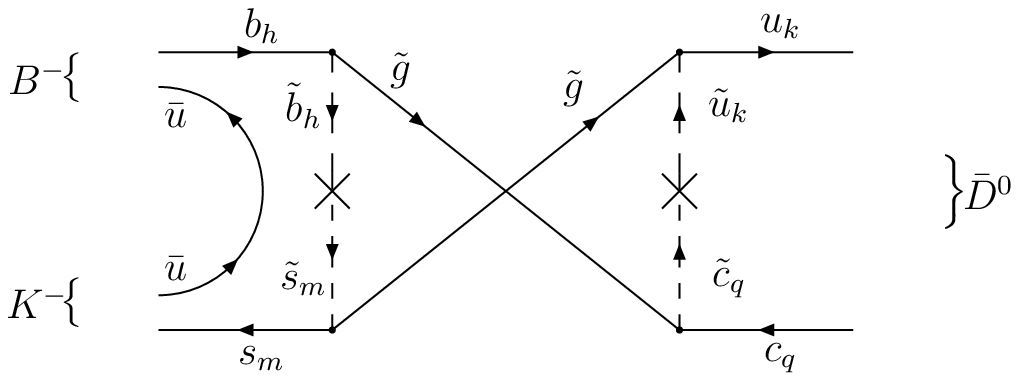, width=8cm, height=3.5cm, angle=0}
\caption{Box diagrams for $B^- \to K^- \bar{D}^0$ ($b\to u
(\bar{c}s)$ transition) with gluino exchanges, where
$h,k,m,n=\{L,R\}$.} \label{gluinofig}
\end{figure}
\begin{figure}[h]
\epsfig{file=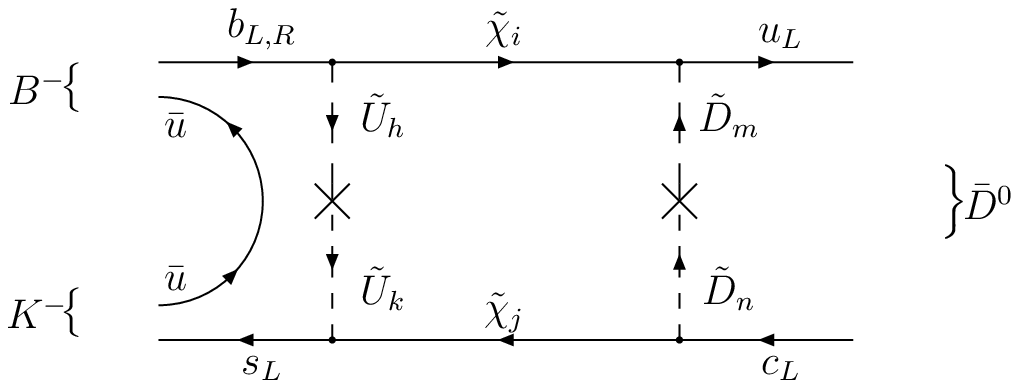, width=8cm, height=3.5cm, angle=0}
\hspace{0.5cm}\epsfig{file=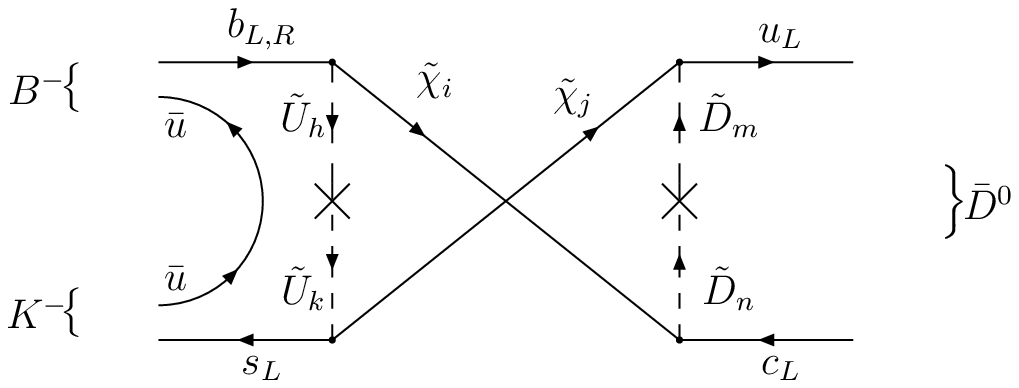, width=8cm, height=3.5cm,
angle=0} \caption{Box diagrams for $B^- \to K^- \bar{D}^0$ ($b\to u
(\bar{c}s)$ transition) with chargino exchanges, where
$U=\{u,c,t\}$, $D=\{d,s,b\}$ and $h,k,m,n=\{L,R\}$.} \label{chifig}
\end{figure}
The dominant SUSY contribution to the $b \to u (\bar{c}s)$
transition can be generated through the box-diagrams with gluino
exchange, as in Fig.\ref{gluinofig}, and chargino exchange, as in
Fig.\ref{chifig}. From these figures, one can see that the $b\to u
(\bar{c}s)$ transition is based on two topologically distinct box
diagrams only for gluino or chargino exchange. This is unlike the
$b\to d$ and $b\to s$ transitions that contribute to $B-\bar{B}$
mixing, where four topologically distinct box diagrams are
included \cite{Gabrielli:2002fr}. Therefore, it is expected that
the Wilson coefficients for this process are different from those
obtained in the literature for $b\to s$ transition. It is also
worth mentioning that contributions through penguin diagrams to
these transitions are always hybrid ({\it i.e.,} contain internal
SUSY and SM particles). Therefore, they are suppressed by $V_{ub}$
in addition to the usual loop suppression factor, hence they are
much smaller than the pure SM or pure SUSY contributions. Thus,
the Wilson coefficients at $m_W$ scale can be expressed as follows
\be C_i^u = (C_i^u)^{SM} + (C_i^u)^{\tilde{g}} +
(C_i^u)^{\tilde{\chi}}, \ee

We evaluate the SUSY contributions to the Wilson coefficients by
using the mass insertion approximation. The Mass insertion
approximation is quite useful method in order to perform model
independent analysis of flavor changing processes in general SUSY
models. In our analysis we set to zero the contributions that are
proportional to to the Yukawa coupling of light quarks. Also, we
use the approximation of retaining only terms proportional to
order $\lambda$. In the case of the gluino exchange all the above
operators give significant contributions and the corresponding
Wilson coefficients are given by

\bea C_1^{\tilde{g}}(m_W) &=&
\frac{\alpha_s^2}{48\tilde{m}^2}(\delta^d_{LL})_{23}
(\delta^u_{LL})_{12} \left[ 7 \tilde{f}_6(x) -4 x f_6(x)
\right],\label{C1}\\
C_2^{\tilde{g}}(m_W) &=&
\frac{\alpha_s^2}{144\tilde{m}^2}(\delta^d_{LL})_{23}
(\delta^u_{LL})_{12} \left[ \tilde{f}_6(x) + 20 x f_6(x)
\right],\label{C2}\\
C_3^{\tilde{g}}(m_W) &=&
\frac{5\alpha_s^2}{48\tilde{m}^2}(\delta^d_{RL})_{23}
(\delta^u_{LR})_{12} \tilde{f}_6(x),\\
C_4^{\tilde{g}}(m_W) &=&
\frac{11\alpha_s^2}{144\tilde{m}^2}(\delta^d_{RL})_{23}
(\delta^u_{LR})_{12}  \tilde{f}_6(x) ,\\
C_5^{\tilde{g}}(m_W) &=& \frac{2
\alpha_s^2}{3\tilde{m}^2}(\delta^d_{RL})_{23}
(\delta^u_{RL})_{12} x f_6(x), \\
C_6^{\tilde{g}}(m_W) &=&
\frac{-\alpha_s^2}{9\tilde{m}^2}(\delta^d_{RL})_{23}
(\delta^u_{RL})_{12} x f_6(x) ,\\
C_7^{\tilde{g}}(m_W) &=&
\frac{\alpha_s^2}{12\tilde{m}^2}(\delta^d_{RR})_{23}
(\delta^u_{LL})_{12}\left[ -\tilde{f}_6(x) + 7 x f_6(x)\right] ,\\
C_8^{\tilde{g}}(m_W) &=&
\frac{\alpha_s^2}{36\tilde{m}^2}(\delta^d_{RR})_{23}
(\delta^u_{LL})_{12}\left[5 \tilde{f}_6(x)+ x f_6(x) \right], \\
C_9^{\tilde{g}}(m_W) &=&
-\frac{\alpha_s^2}{48\tilde{m}^2}(\delta^d_{RL})_{23}
(\delta^u_{RL})_{12} x f_6(x) ,\\
C_{10}^{\tilde{g}}(m_W) &=&
\frac{5\alpha_s^2}{144\tilde{m}^2}(\delta^d_{RL})_{23}
(\delta^u_{RL})_{12} x f_6(x).\eea where $x =
m_{\tilde{g}}^2/\tilde{m}^2$. The $m_{\tilde{g}}$ is the gluino
mass and the $\tilde{m}^2$ is an average squark mass. The
functions $f_6(x)$ and  $\tilde{f}_6(x)$ are the same as the loop
function obtained in case of $b\to d (\bar{q}q)$ and are given by
\bea f_6(x) &=& \frac{6(1+3 x) \ln x + x^3 - 9 x^2 - 9 x +
17}{6(x-1)^5},\label{f6}\\
\tilde{f}_6(x) &=& \frac{6x(1+x) \ln x - x^3 - 9 x^2 + 9 x +
1}{3(x-1)^5}. \label{f6tilde}\eea The Wilson coefficients
$\tilde{C}_i^{\tilde{g}}$ are simply obtained by interchanging $L
\leftrightarrow R$ in the mass insertions appearing in
$C_i^{\tilde{g}}$. The above Wilson coefficients are due to the
gluino exchange of $b\to u$ transition, the corresponding
coefficients for $b\to c$ transition can be obtained by changing
the mass insertions $(\delta^u_{AB})_{12}$ to
$(\delta^u_{AB})_{21}$ where $\{A,B\}=\{L,R\}$.

Note that the discrepancy between the above SUSY Wilson
coefficients of $b\to u$ transition and those of $b\to d$ or $b\to
s$ $\Delta B=2$ transition is due to the following reasons: 1) the
$b\to u$ transition is based, as mentioned above, on two distinct
box diagrams only in contrast of the $\Delta B=2$ transition where
four distinct box diagrams are involved. 2) All the external
quarks in the box diagrams of $b\to u(\bar{c}s)$ are different,
therefore, one can not use Fierz transformation to relate any
operator with the other unlike the case in $\Delta B=2$. For
instance, in $B_d - \bar{B}_d$ mixing the operate $Q_2 =
\left(\bar{d}^{\alpha} \gamma_{\mu} L
b^{\beta}\right)\left(\bar{d}^{\beta} \gamma_{\mu} L
b^{\alpha}\right)$ is equivalent to the operator
$Q_2=\left(\bar{d}^{\alpha} \gamma_{\mu} L
b^{\alpha}\right)\left(\bar{d}^{\beta} \gamma_{\mu} L
b^{\beta}\right)$. In this case, the Wilson coefficients $C_1$ and
$C_2$ in Eqs.(\ref{C1},\ref{C2}) are combined together and leads
to to the usual $\Delta B=2$ Wilson coefficient: $C_1 \propto
\alpha_s/(108 \tilde{m}^2) \left(24 x f_6(x) + 66
\tilde{f}_6(x)\right)$ \cite{Gabrielli:2002fr}. In this respect,
it is clear that the expression used in Eq.(9) in
Ref.\cite{Chang:2001ah} for $H_{\rm{eff}}^{b\to u(\bar{c}s)}$ is
incorrect.

Now let us turn to the chargino contributions to the effective
Hamiltonian in Eq.(\ref{effH}) in the mass insertion
approximation. The leading diagrams are illustrated in
Fig.\ref{chifig}, where the cross in the middle of the squark
propagator represents a single mass insertion. Within the above
mentioned approximation where we neglect contributions
proportional to the light quark masses, one find that the relevant
chargino exchange affects only the operator $Q_1$, as in the SM
and the corresponding Wilson coefficient is given by \bea
C_1^{\tilde{\chi}}(m_W)\!&\!=\!&\! \frac{\alpha \sqrt{\alpha}}{16
\tilde{m}^2}\left[ \sqrt{\alpha} V^*_{i1} V_{j1} U_{i1}
U^*_{j1}(\delta^d_{LL})_{12}\left( (\delta^u_{LL})_{23} + \lambda
(\delta^u_{LL})_{13}\right) \right.\nonumber\\
\!&\!+\!&\!\left. \frac{y_t}{\sqrt{4\pi}} U_{i1} U^*_{j1} V_{j1}
V^*_{i2}(\delta^d_{LL})_{12} \left( (\delta^u_{LR})_{23} + \lambda
(\delta^u_{LR})_{13}
\right)\right]\left(L_2(x_i,x_j)-2L_0(x_i,x_j)\right),~~~~~\eea
where $\alpha=g^2/4\pi$ and $g$ is the $SU(2)$ gauge coupling
constant. The $\lambda$ parameter stands for the Cabibbo mixing,
\ie, $\lambda =0.22$. The $U_{ij}$ and $V_{ij}$ are the unitary
matrices that diagonalise the chargino mass matrix and $y_t$ is
the top yukawa coupling. The $x_i =
m^2_{\tilde{\chi}_i}/\tilde{m}^2$, and the functions $L_0(x, y)$
and $L_2(x, y)$ are given by \cite{Gabrielli:2002fr} \bea L_0(x,y)
&=& \sqrt{xy} ~\frac{x h_0(x) - y h_0(y)}{x-y}, ~~~~~h_0(x) =
\frac{-11+7 x-2
x^2}{(1-x)^3} - \frac{6\ln x}{(1-x)^4},\nonumber\\
L2(x,y)&=& \frac{x h_2(x)- y h_2(y)}{x-y}, ~~~~~~~~~~~~~~~~h_2(x)
= \frac{2+3 x-x^2}{(1-x)^3} + \frac{6x \ln x}{(1-x)^4}.
\label{L2}\eea Finally, we have also neglected the small
contributions from the box diagrams where both gluino and chargino
are exchanged as in Ref.\cite{Chakraverty:2003uv}.

To obtain the Wilson-coefficients at the scale $m_b$ one has to
solve the corresponding renormalisation group equations. The
solution is generally expressed as \be C_i(m_b) = \sum_{j}
U_{ij}(m_b,m_W) C_j(m_W), \ee where $U_{ij}(m_b,m_W)$ is the
evolution matrix given by the $8\times 8$ anomalous dimension
matrix of leading order (LO) corrections in QCD
\cite{Buchalla:1995vs}. Note that we have not included the
operators $Q_{9,10}$ since they have zero matrix elements at LO
and also they do not mix with the other operators in the evolution
from $m_W$ scale down to $m_b$ scale.
\begin{equation}
U(m_b,m_W)= \hat{V}\left(\left[\frac{\alpha_s(m_W)}{\alpha_s(m_b
)}\right]^{-\frac{\vec{\gamma^{(0)}}}{2\beta_0}}\right)_D\hat{V}^{-1}
\end{equation}
where $\hat{V}$ diagonalizes the $\hat{\gamma}^{(0)T}$
\begin{equation}
\hat{\gamma}^{(0)}_D =\hat{V}^{-1}\hat{\gamma}^{(0)T}\hat{V}
\end{equation}
and $\vec{\gamma}^{(0)}$ is the diagonal elements of
$\hat{\gamma}^{(0)}_D$. The value of $\beta_0$ is given by
$\beta_0= \frac{11}{3} N_c - \frac{2}{3} f$ where $N_c$ is the
number of colors and $f$ is number of active flavors. Finally, the
anomalous dimension matrix $\hat{\gamma}^0$ at the leading order
is given by \be \hat{\gamma}^{0}=
\left(\begin{array}{cccccccc} -2 & 6 & 0&0 &0 &0 &0 &0\\
6 & -2 &0 & 0& 0&0 &0 &0 \\
0 &0 & 2 & -6 & 0& 0&0 &0\\
 0& 0& 0 & -16& 0& 0& 0&0\\
 0& 0& 0& 0& -2 & 6 & 0& 0\\
0 & 0& 0& 0& 6 & -2 & 0&0 \\
 0& 0& 0& 0& 0& 0& 2 & -6\\
0 & 0& 0& 0& 0& 0& 0 & -16
\end{array}\right).
\ee As can be seen from the above matrix that the mixing between
different operators is divided into blocks. Each block contains
two operators $(Q_i,Q_{i+1})$, $i=1,3,4,7$ and with no mixing
between different blocks \cite{Buras:1992tc}.

Let us now consider the evaluation of the hadronic matrix elements
of the above operators which represents the most uncertain part in
this calculation. In the limit of neglecting QCD corrections and
$m_b \gg \Lambda_{QCD}$, the hadronic matrix elements of $B^-\to D
K^-$ decay can be factorized. The hadronic matrix elements for the
operators $Q_i^u$ are given by \be \langle D^0 K^- \vert~ Q_i^u
~\vert B^- \rangle =0,\ee and \bea \langle \bar{D}^0 K^- \vert~
Q_1^u ~\vert B^- \rangle &=& - \frac{X}{3},
\nonumber\\
\langle \bar{D}^0  K^- \vert~ Q_2^u ~\vert B^- \rangle &=&- X,\nonumber\\
\langle \bar{D}^0  K^- \vert~ Q_3^u ~\vert B^- \rangle &=&
\frac{2m_D^2}{3(m_b-m_s)(m_u+m_c)} X,\nonumber\\
\langle \bar{D}^0 K^- \vert~ Q_4^u ~\vert B^- \rangle &=&
\frac{2 m_D^2}{(m_b-m_s)(m_u+m_c)} X,\nonumber\\
\langle \bar{D}^0  K^- \vert~ Q_5^u ~\vert B^- \rangle &=&
\langle \bar{D}^0  K^- \vert~ Q_6^u ~\vert B^- \rangle =0,\nonumber\\
\langle \bar{D}^0 K^-\vert~ Q_7^u ~\vert B^- \rangle &=&\frac{X}{6} \\
\langle \bar{D}^0 K^-\vert~ Q_8^u ~\vert B^- \rangle
&=&\frac{X}{2},
\nonumber\\
\langle \bar{D}^0 K^-\vert~ Q_9^u ~\vert B^- \rangle &=& \langle
\bar{D}^0 K^-\vert Q_{10}^u \vert~ B^- ~\rangle =0.\nonumber \eea
While the hadronic matrix elements for the operators $Q_i^c$ are
given as follows: \be \langle \bar{D}^0 K^- \vert~ Q_i^c ~\vert
B^- \rangle =0,\ee and \bea \langle D^0 K^- \vert~ Q_1^c ~\vert
B^- \rangle
&=&- Y -\frac{1}{3}X,\nonumber\\
\langle D^0 K^- \vert~ Q_2^c ~\vert B^- \rangle &=& -\frac{1}{3}Y-
X, \nonumber\\
\langle D^0 K^- \vert~ Q_3^c ~\vert B^- \rangle &=& Y +
\frac{2m_D^2}{3(m_b-m_s)(m_u+m_c)} X,\nonumber\\
\langle D^0 K^- \vert~ Q_4^c ~\vert B^- \rangle &=&\frac{1}{3} Y +
\frac{2 m_D^2}{(m_b-m_s)(m_u+m_c)} X,\nonumber\\
\langle D^0 K^- \vert~ Q_5^c ~\vert B^- \rangle &=& \langle D^0
K^-
\vert Q_6^c \vert~ B^- ~\rangle = 0,\nonumber\\
\langle D^0 K^-\vert~ Q_7^u ~\vert B^- \rangle &=&-\frac{m_K^2}{(m_b-m_c)(m_u+m_s)} Y + \frac{1}{6}X \nonumber\\
\langle D^0 K^-\vert~ Q_8^c ~\vert B^- \rangle
&=&-\frac{1}{3}\frac{m_K^2}{(m_b-m_c)(m_u+m_s)} Y + \frac{1}{2}X
,\nonumber\\
\langle D^0 K^-\vert~ Q_9^c ~\vert B^- \rangle &=& \langle D^0
K^-\vert~ Q_{10}^c ~\vert B^- \rangle =0, \eea where $X$ and $Y$ are
given in Eq.(\ref{XY}).

Having evaluated the SUSY contributions to the Wilson coefficients
and determined the hadronic matrix elements of the relevant
operators, one can analyze the decay amplitude of $B^- \to D K^-$
and study the SUSY effect on the CP asymmetries $A_{\pm}$ and
branching ratios $R_{\pm}$. As can be observed, the Wilson
coefficients depends on several mass insertions which are in
general complex and provide new sources for the CP violation
beyond the SM phase in the CKM mixing matrix. These new CP
violating phases may contribute significantly to the $b\to u$
transition and affect the determination of the angle $\gamma$.
Nevertheless, one should be very careful with the constraints
imposed on these parameters. In fact, the dominant gluino
contributions depend on the mass insertions:
$(\delta^u_{AB})_{12}$ and $(\delta^d_{AB})_{23}$. The mass
insertions $(\delta^d_{AB})_{23}$ are constrained by the
experimental results for the branching ratio of $B\to X_s \gamma$
\cite{Khalil:2005qg,Besmer:2001cj,Ciuchini:2002uv}. These
constraints are very weak on the $LL$ or $RR$ mass insertion and
more stronger for $LR$ or $RL$ mass insertions. Concerning the
mass insertion $(\delta^u_{AB})_{12}$, the chargino contributions
to the $K^0-\bar{K}^0$ impose constraint on the $LL$ mass
insertion only \cite{Khalil:2001wr}. However, the $D^0-\bar{D}^0$
mixing may induce more strangest constraints on both $LL(RR)$ and
$LR(RL)$ mass insertions. Therefore, before we proceed in
analyzing the decay amplitude of $B^- \to D K^-$,  we will take a
short detour and give a detail analysis for the SUSY contributions
to $D^0-\bar{D}^0$ mixing and revise the corresponding constraints
on the $(\delta^u_{AB})_{12}$ mass insertions.

%
\section{{\large \bf Constraints from $D^0-\bar{D}^0$ mixing}}

We start this section by summarizing the SM results for the
$D^0-\bar{D}^0$ mixing, then we consider the supersymmetric
contribution to the effective Hamiltonian for $\Delta C=2$
transitions given by gluino and chargino box exchanges.

In the $D^0$ and $\bar{D}^0$ systems, the flavor eigenstates are
given by $D^0= (\bar{u} c)$ and $\bar{D}^0= (u \bar{c})$. The
formulism for $D^0-\bar{D}^0$ mixing is the same as the one used
for $K^0-\bar{K}^0$ and $B^0-\bar{B}^0$ mixing. The mass
eigenstates are given in terms of the weak eigenstates as: \be
D_{1,2} = p~ D^0 \pm q~ \bar{D}^0, \ee where the ratio $q/p$ can
be written in terms of the off-diagonal element of the mass
matrix: $q/p = \sqrt{M_{12}^*/M_{12}}$ and $q/p \neq 1$ is an
indication for the CP violation through mixing. The strength of
$D^0-\bar{D}^0$ mixing is described by the mass difference $$
\Delta M_D = M_{D_1} - M_{D_2}.$$ The present experimental results
for $\Delta M_D $ is given by \cite{Cawlfield:2005ze} \be \left(
\Delta M_{D} \right)_{\rm{exp}} < 1.7 \times 10^{-13}~ {\rm GeV}.
\ee

The CP asymmetry of the $D^0$ and $\bar{D}^0$ meson decay to CP
eigenstate $f$ is given by \bea a_f(t) &=& \frac{\Gamma(D^0\to f)
- \Gamma(\bar{D}^0 \to f)}{\Gamma(D^0\to f) + \Gamma(\bar{D}^0 \to
f)}
\nonumber\\
&=& S_f \sin(\Delta M_D t) + C_f \cos(\Delta M_D t), \eea where
$S_f$ and $C_f$ represent the mixing and direct CP asymmetry
respectively and they are given by \be S_f = \frac{2 {\rm
Im}\left[\frac{q}{p} \bar{\rho}(f)\right]}{\vert
\bar{\rho}(f)\vert^2 +1}, ~~~~~~~ C_f = \frac{\vert
\bar{\rho}(f)\vert^2 -1}{\vert \bar{\rho}(f)\vert^2 +1}. \ee The
parameter $\bar{\rho}(f)$ is defined by $\bar{\rho}(f) =
\frac{{\cal A}(\bar{D}^0 \to f)}{{\cal A}(D^0 \to f)}$.
Generically, the $\Delta M_D$ and $S_f$ can be calculated by \be
\Delta M_D = 2\left\vert \langle D^0 \vert H_{eff}^{\Delta C=2}
\vert \bar{D}^0 \rangle \right\vert,~~~~~ S_f = \sin\left({\rm
arg}\left[\langle D^0 \vert H_{eff}^{\Delta C=2} \vert \bar{D}^0
\rangle\right]\right) . \ee Here $H_{eff}^{\Delta C=2}$ is the
effective Hamiltonian responsible for $\Delta C=2$ transition. In
the framework of the SM, this transition occurs via box diagram in
which two virtual down quarks and two virtual W bosons are
exchanged. The $\Delta M^{SM}_D = 2 \vert \langle D^0 \vert
(H_{eff}^{\Delta C=2})^{SM} \vert \bar{D}^0 \rangle \vert$ is
given by \cite{Datta:1984jx} \be  \Delta M^{SM}_D \simeq
\frac{G_F^2}{3 \pi^2} M_D^2 f_D^2 \vert V^*_{cs}V_{cd} \vert^2
\frac{(m_s^2 -m_d^2)^2}{m_c^2} \simeq 1.4 \times 10^{-18}~GeV. \ee
As can be seen from this expression, the SM predicts a very small
$D^0-\bar{D}^0$ mixing. Note that, in the above estimation for
$\Delta M^{SM}_D $, the $b$-quark contribution has been neglected
since it is much smaller due to the CKM suppression. Also, the CP
violation is absent in the mixing and in the dominant tree level
decay due to the involving of the first two generations only.

In supersymmetric theories, the dominant contributions to the off
diagonal entry in the $D^0$-meson mass matrix, $\mathcal{M}_{12} =
\langle D^0 \vert H_{\mathrm{eff}}^{\Delta C=2} \vert \bar{D}^0
\rangle$, is given by
\begin{equation}
\mathcal{M}_{12} = \mathcal{M}_{12}^{\mathrm{SM}}+
\mathcal{M}_{12}^{\tilde{g}} + \mathcal{M}_{12}^{\tilde{\chi}^+},
\end{equation}
where $\mathcal{M}_{12}^{\tilde{g}}$, and
$\mathcal{M}_{12}^{\tilde{\chi}^+}$ correspond to the gluino and
chargino contributions respectively. The effect of supersymmetry
can be parameterized as follows
\begin{equation}
r_c^2 e^{2 i \theta_c} =
\frac{\mathcal{M}_{12}}{\mathcal{M}_{12}^{\mathrm{SM}}},
\label{parametrization}
\end{equation}
where $\Delta M_{D} = 2 \vert \mathcal{M}_{12}^{\mathrm{SM}}\vert
r_c^2$ and $2 \theta_c =
\mathrm{arg}\left(1+\frac{\mathcal{M}_{12}^{\mathrm{SUSY}}}
{\mathcal{M}_{12}^{\mathrm{SM}}}\right)$. As in the case of $K^0$
and $B^0$ systems, The most general effective Hamiltonian for
$\Delta C=2$ processes, induced by gluino and chargino exchanges
through box diagrams, can be expressed as \be H^{\Delta C=2}_{\rm
eff}=\sum_{i=1}^5 C_i(\mu) Q_i(\mu) + \sum_{i=1}^3
\tilde{C}_i(\mu) \tilde{Q}_i(\mu) + h.c. \;, \label{Heff} \ee
where $C_i(\mu)$, $\tilde{C}_i(\mu)$ and $Q_i(\mu)$,
$\tilde{Q}_i(\mu)$ are the Wilson coefficients and operators
respectively renormalized at the scale $\mu$, with \bea Q_1
&=&\bar{u}_L^{\alpha} \gamma_{\mu} c_L^{\alpha}~ \bar{u}_L^{\beta}
\gamma_{\mu} c_L^{\beta},~~~~ Q_2 = \bar{u}_R^{\alpha}
c_L^{\alpha}~ \bar{u}_R^{\beta} c_L^{\beta},~~~~ Q_3 =
\bar{u}_R^{\alpha} c_L^{\beta}~
\bar{u}_R^{\beta} c_L^{\alpha},\nonumber\\
Q_4 &=& \bar{u}_R^{\alpha} c_L^{\alpha}~ \bar{u}_L^{\beta}
c_R^{\beta},~~~~~~~~~ Q_5 = \bar{u}_R^{\alpha} c_L^{\beta}~
\bar{u}_L^{\beta} c_R^{\alpha}\, . \label{operators} \eea In
addition, the operators $\tilde{Q}_{1,2,3}$ are obtained from
$Q_{1,2,3}$ by exchanging $L \leftrightarrow R$.

In the case of the gluino exchange all the above operators give
significant contributions and the corresponding Wilson coefficients
are given by \cite{gluino}\bea C_1^{\tilde{g}}(m_W) &=&
-\frac{\alpha_s^2}{216 m_{\tilde{q}}^2} \left(24x f_6(x)+
66\tilde{f}_6(x)
\right) (\delta_{12}^u)^2_{LL},\nonumber\\
C_2^{\tilde{g}}(m_W) &=& -\frac{\alpha_s^2}{216 m_{\tilde{q}}^2} 204
x  f_6(x)
(\delta_{12}^u)^2_{RL},\nonumber\\
C_3^{\tilde{g}}(m_W) &=& \frac{\alpha_s^2}{216 m_{\tilde{q}}^2} 36 x
f_6(x)
(\delta_{12}^u)^2_{RL},\\
C_4^{\tilde{g}}(m_W) &=& -\frac{\alpha_s^2}{216 m_{\tilde{q}}^2}
\left[\left(504 x f_6(x) - 72 \tilde{f}_6(x)\right)
(\delta_{12}^u)_{LL}(\delta_{12}^u)_{RR} -132 \tilde{f}_6(x)
(\delta_{12}^u)_{LR}(\delta_{12}^u)_{RL}\right] ,
\label{wilsong}\nonumber\\
C_5^{\tilde{g}}(m_W) &=& -\frac{\alpha_s^2}{216 m_{\tilde{q}}^2}
\left[\left(24 x f_6(x) +120 \tilde{f}_6(x)\right)
(\delta_{12}^u)_{LL}(\delta_{12}^u)_{RR} -180 \tilde{f}_6(x)
(\delta_{12}^u)_{LR}(\delta_{12}^u)_{RL}\right],\nonumber \eea where
$x=m^2_{\tilde{g}}/\tilde{m}^2$ and $\tilde{m}^2$ is an average
squark mass. The functions $f_6(x)$ and $\tilde{f}_6(x)$ are given
in Eqs.(\ref{f6},\ref{f6tilde}). The Wilson coefficients
$\tilde{C}_{1-3}$ are simply obtained by interchanging
$L\leftrightarrow R$ in the mass insertions appearing in $C_{1-3}$.

In the case of the chargino exchange the operator $Q_1$ only gives a
significant contribution \cite{Khalil:2001wr}. At the first order in
the mass insertion approximation, the Wilson coefficient
$C_{1}^{\chi}(m_W)$ is give by \be C_1^{\chi}(m_W) = \frac{g^4}{768
\pi^2 \tilde{m}^2}~ \sum_{i,j}~ \vert V_{i1}\vert^2~ \vert V_{j1}
\vert^2~ (\delta^d_{21})_{LL}^2 L_2(x_i,x_j). \ee where $x_i=
m^2_{\tilde{\chi}_i^+}/\tilde{m}^2$, and the function $L_2(x,y)$ is
as given in Eq.(\ref{L2}).

As usual, the Wilson coefficients $C_i(\mu)$ are related to
$C_i(m_W)$ by \cite{gluinoB} \be C_r(\mu) = \sum_i \sum_s \left(
b_i^{(r,s)} + \eta c_{i}^{(r,s)}\right) \eta^{a_i} C_s(m_W),
\label{CWnlo} \ee where $\eta=\alpha_S(m_W)/\alpha_S(\mu)$ and the
coefficients $b_i^{(r,s)}$, $c_i^{(r,s)}$, and $a_i$ appearing in
(\ref{CWnlo}) can be found in Ref.\cite{gluinoB}. Also the matrix
elements of the operators $Q_i$ in the vacuum insertion
approximation are given by \cite{gluino} \bea \langle D^0 \vert
Q_1 \vert \bar{D}^0 \rangle &=& \frac{1}{3} M_{D} f_{D}^2,\nonumber\\
\langle D^0 \vert Q_2 \vert \bar{D}^0 \rangle &=& -\frac{5}{24}
\left(\frac{M_{D}}
{m_c + m_u} \right)^2 M_{D} f_{D}^2 ,\nonumber\\
\langle D^0 \vert Q_3 \vert \bar{D}^0 \rangle &=& \frac{1}{24}
\left(\frac{mM_{D}} {m_c + m_u} \right)^2 M_{D} f_{D}^2 ,
\label{matrixelements}
\\
\langle D^0 \vert Q_4 \vert \bar{D}^0 \rangle &=&
\left[\frac{1}{24} +\frac{1}{4}\left(\frac{M_{D}}
{m_c + m_u} \right)^2 \right] M_{D} f_{D}^2,\nonumber\\
\langle D^0 \vert Q_5 \vert \bar{D}^0 \rangle &=&
\left[\frac{1}{8} + \frac{1}{12} \left(\frac{M_{D}}{m_c + m_u}
\right)^2\right] M_{D} f_{D}^2.\nonumber \eea The same results are
also valid for the corresponding operators $\tilde{Q}_i$ since
strong interactions preserve parity.

We now discuss the results of SUSY contribution to $D^0-\bar{D}^0$
mixing. It is worth mentioning that the mass insertions
$(\delta^d_{AB})_{12}$ are strongly constrained by the experimental
limits of $K^0 - \bar{K}^0$ mixing. In particular, the $\Delta M_K$
upper bound implies that $\vert (\delta^d_{LL})_{12}^2 \vert \leq
10^{-4}$ \cite{gluino}. Therefore, the chargino contribution to
$\Delta M_D$ becomes very suppressed and can be neglected with
respect to the gluino contributions which depend on less constrained
mass insertions $(\delta^u_{AB})_{12}$ . As an example, we present
the gluino contribution to $\Delta M_D$, with $m_{\tilde{g}} \simeq
m_{\tilde{q}} \simeq 500$ GeV: \bea \frac{\Delta
M_D^{SUSY}}{1.7\times 10^{-13}} &\simeq& \Big\vert~ 33.4
(\delta^u_{LL})^2_{12} + 1733.6 (\delta^u_{LR})^2_{12} - 3178.5
(\delta^u_{LR})_{12}(\delta^u_{RL})_{12} +1733.6 (\delta^u_{RL})^2_{12} \nonumber\\
&-& 12946.9 (\delta^u_{LL})_{12}(\delta^u_{RR})_{12} + 33.4
(\delta^u_{RR})^2_{12} ~\Big \vert < 1. \eea From this expression,
we can see that the strongest constraint will be imposed on the
product $(\delta^u_{LL})_{12}(\delta^u_{RR})_{12}$ while the
constraint obtained on the individual mass insertion
$(\delta^u_{LL})_{12}$ or $(\delta^u_{RR})_{12}$ is less
stringent.

As usual in this kind of analysis, the most conservative constraints
on the mass insertions can be obtained by considering the
contribution due to a single mass insertion per time and set all
other ones to zero. In table I, we present the results for the upper
bounds on the relevant mass insertions from the experimental
constraint on $\Delta M_D$ for $x=1/4, 1,$ and $4$. We find that
these bounds on $(\delta^u_{AB})_{12}$ are more stringent than those
obtained from the chargino contribution to the $K^0- \bar{K}^0$ in
Ref.\cite{Khalil:2001wr}. In fact, the $(\delta^u_{LR})_{12}$ and
$(\delta^u_{RL})_{12}$ are completely unconstrained by the chargino
contribution to $K^0 -\bar{K}^0$ mixing. Therefore, their bounds in
the above table are the only known constraints. However, we should
mention that these constraints may be relaxed if one consider
simultaneous contributions from more than one mass insertion. In
this case, a possible cancellation may occur which reduce the SUSY
contribution significantly and leave a room for a larger mass
insertion.
\begin{table}
\begin{center}
\begin{tabular}{|c|c|c|c|c|}
  \hline
  x & $\sqrt{\vert(\delta^u_{LL(RR)})_{12}^2\vert}$ & $\sqrt{\vert(\delta^u_{LR(RL)})_{12}^2\vert}$ &
  $\sqrt{\vert(\delta^u_{LL})_{12} (\delta^u_{RR})_{12}\vert}$ &
  $\sqrt{\vert(\delta^u_{LR})_{12} (\delta^u_{RL})_{12}\vert}$ \\
  \hline
  1/4 & $7.5\times 10^{-2}$ & $2.1\times 10^{-2}$ & $7.5\times 10^{-3}$  & $1\times 10^{-2}$  \\
  1 & $1.7\times 10^{-2}$  & $2.4\times 10^{-2}$  & $8.7\times 10^{-3}$  & $1.7\times 10^{-2}$  \\
  4 & 0.4 & $3.3\times 10^{-2}$  & $1.2\times 10^{-2}$  & $4\times 10^{-2}$  \\
  \hline
\end{tabular}
\caption{Upper bounds on
$\sqrt{\vert(\delta^u_{AB)})_{12}^2\vert}$ from $\Delta M_D < 1.7
\times 10^{-13}$ GeV, for $x=m_{\tilde{g}}^2/m_{\tilde{q}}^2= 1/4,
1, 4$.}
\end{center}
\end{table}

Finally, we comment on the CP violation in this process. As
emphasized above, the SM contribution to $D^0- \bar{D}^0$ is real
since it is proportional to $V_{cs}^*V_{cd}$. Furthermore, it is
much smaller than the dominant gluino contribution. Therefore CP
violating phase $\theta_c$ in Eq.(\ref{parametrization}) can be
written as \be \theta_c = \frac{1}{2}~ \rm{arg}\left(\frac{{\cal
M}_{12}}{{\cal M}_{12}^{\rm{SM}}} \right) \simeq ~ \frac{1}{2}~
\rm{arg}\left({\cal M}_{12}^{\tilde{g}}\right). \ee In case
$(\delta^u_{LR})_{12}$ gives a dominant contribution to ${\cal
M}_{12}^{\tilde{g}}$, $\theta_c$ will be given by \be \theta_c =
\frac{1}{2}~ \rm{arg}\left((\delta^u_{LR})_{12}\right) \simeq
{\cal O}(1),\label{thetac}\ee which means that the mixing CP
asymmetry of $D^0- \bar{D}^0$, $S_f$, could be quite large.
Therefore, one can conclude that the new physics in general and
supersymmetry in particular could enhance the $D^0- \bar{D}^0$
mixing significantly.

%
%
\section{{\large \bf Supersymmetric contribution to $R_{\pm}$ and $A_{\pm}$}}

In this section we study the supersymmetric contributions to the
CP asymmetries and the branching ratios of $B^- \to D K^-$ decay
in the following cases: 1) negligible $D^0-\bar{D}^0$ mixing. 2)
Large $D^0-\bar{D}^0$ mixing due to a possible significant SUSY
contribution as advocated in the previous section.

In general, applying the naive factorization approximation implies
that the amplitudes $ A(B^- \to D K^-)$ are given by \be A(B^- \to
D^0 K^-)= \sum_{i=1}^8 \left(C^c_i - \tilde{C}^c_i \right) \langle
D^0 K^- \vert Q^c_i \vert B^- \rangle, \ee and \be A(B^- \to
\bar{D}^0 K^-)= \sum_{i=1}^8 \left(C^u_i - \tilde{C}^u_i \right)
\langle \bar{D}^0 K^- \vert Q^u_i \vert B^- \rangle. \ee The sign
difference between the Wilson coefficients $C_i$ and $\tilde{C}_i$
in the above equations is due to the fact that the initial and
final states of $B^- \to D K^-$  decays have opposite parity and
therefore $\langle D K^- \vert~ Q_i ~\vert B^- \rangle = - \langle
D K^- \vert~ \tilde{Q}_i ~\vert B^- \rangle$ \cite{Khalil:2003bi}.

\subsection{{\large \bf $R_{\pm}$ and $A_{\pm}$ with negligible $D^0 -\bar{D}^0$ mixing}}

In case of neglecting the effect of $D^0 -\bar{D}^0$ mixing, it is
useful to parameterize the SUSY contribution by introducing the
ratio of the SM and SUSY amplitudes as follows: \be
\frac{A^{SUSY}(B^- \to \bar{D}^0 K^-)}{A^{SM}(B^- \to \bar{D}^0
K^-)} = R_1~ e^{i(\phi_1 - \gamma)}~ e^{i \delta_1}, \ee and \be
\frac{A^{SUSY}(B^- \to D^0 K^-)}{A^{SM}(B^- \to D^0 K^-)} = R_2~
e^{i\phi_2}~ e^{i \delta_2}, \ee where $R_i$ stands for the
corresponding absolute value of $\vert A^{SUSY}/A^{SM} \vert$, the
angles $\phi_i$ are the corresponding SUSY CP violating phase, and
$\delta_i=\delta^{SM}_i - \delta^{SUSY}_i$ are the strong phases.
In this respect, our previous definition for the SM ratio of the
amplitudes of $B^- \to \bar{D}^0 K^-$ and $B^- \to D^0 K^-$ in
Eq.(\ref{ratio}) will be generalized as follows \bea \frac{A(B^-
\to \bar{D}^0 K^-)}{A(B^- \to D^0 K^-)}&=& \frac{A^{SM}(B^- \to
\bar{D}^0 K^-)+ A^{SUSY}(B^- \to \bar{D}^0 K^-)}{A^{SM}(B^- \to
D^0 K^-)+A^{SUSY}(B^- \to D^0 K^-)}\nonumber\\
&=& r_B e^{i\delta_B} \left[ \frac{e^{i \gamma} + R_1
e^{i\phi_1}}{1+ R_2 e^{i \phi_2}}\right] \equiv R_B e^{i\delta_B}
e^{i \phi_B}, \eea where \be R_B = r_B \left\vert \frac{e^{i \gamma}
+ R_1 e^{i\phi_1}}{1+ R_2 e^{i \phi_2}}\right\vert, ~~~ \rm{and}~~~
\phi_B = \rm{arg} \left[\frac{e^{i \gamma} + R_1 e^{i\phi_1}}{1+ R_2
e^{i \phi_2}}\right]. \label{RB}\ee Note that, for simplicity, we
have assumed that the SM and SUSY strong phases are equal. In this
case, the ratios $R_{\pm}$ and the CP asymmetries $A_{\pm}$ take the
form \be R_{\pm} = 1 + R_B^2 \pm 2 R_B \cos\delta_B \cos \phi_B,
\label{RCP2}\ee and \be A_{\pm} = \frac{\pm 2 R_B \sin\delta_B \sin
\phi_B}{1 + R_B^2 \pm 2 R_B \cos\delta_B \cos \phi_B} \label{ACP2}.
\ee

As shown in Eq.(\ref{RB}), the deviation of $R_B$ from the
standard model value $r_B$ is governed by the size of $R_1$ and
$R_2$. Therefore, we start our analysis by discussing the dominant
gluino contributions to $R_1$ and $R_2$. We choose the input
parameters as $\tilde{m}=250$ GeV, $x=1$ we obtain \be R_1 = 0.15
(\delta^d_{LL})_{23}(\delta^u_{LL})_{12} - 0.17
(\delta^d_{RR})_{23}(\delta^u_{LL})_{12} + 0.18
(\delta^d_{RL})_{23}(\delta^u_{LR})_{12} - \{L \leftrightarrow R
\},~~~~~~~ \ee and \bea R_2 &=& - 0.01
(\delta^d_{LL})_{23}(\delta^u_{LL})_{21} - 0.015
(\delta^d_{RR})_{23}(\delta^u_{LL})_{21} + 0.03
(\delta^d_{RL})_{23}(\delta^u_{LR})_{21} - \{L \leftrightarrow R
\}.~~~~~~~   \eea Using the fact that the mass insertion is less
than or equal one, we find that $R_2 \ll 1$, i.e $A^{SUSY}(B^- \to
D^0 K^-) \ll A^{SM}(B^- \to D^0 K)$. It is worth mentioning that
$(\delta^d_{AB})_{23}$ is constrained by the experimental results
for $B \to X_s \gamma$ decay. These constraints are very weak on
the $LL$ and $RR$ mass insertions and they can be of order one.
However, they impose stringent upper bounds on the $LR$ and $RL$
mass insertions, namely $\vert (\delta^d_{LR(RL)})_{23} \vert
\lsim 1.6 \times 10^{-2}$ \cite{Khalil:2005qg}. Concerning the
$(\delta^u_{AB})_{12}$, the important constraints on these mass
insertions are due to the $D^0 -\bar{D}^0$ mixing. Applying these
constraints one finds that $R_1$ is also quite small and the SM
gives the dominant contribution. Therefore, there will be no
chance to modify the results obtained in Fig. 2. However, as
advocated in the previous section, these constraints can be
relaxed if one allows for simultaneous contributions from more
than one mass insertion, which is the case in any realistic model.
In this case, there may be cancellation between different
contributions which reduces the SUSY contribution to $D^0
-\bar{D}^0$ mixing without severely constraining the mass
insertion. If we adopt this scenario and assume, for instance,
that $(\delta^d_{LL})_{23} \simeq -(\delta^d_{RR})_{23}$ and
$(\delta^u_{LL})_{12} \simeq -(\delta^u_{RR})_{12}$, then one can
easily see that $R_1 \simeq {\mathcal{O}(0.6)}$ and the phase
$\phi_1$ is given by
$\mathrm{arg}\left[(\delta^d_{LL})_{23}+(\delta^u_{LL})_{12}\right]$.

In this case, one can easily observe that different combinations
of $(\gamma, \phi_1)$ can lead to values for the $A_{\pm}$ within
the experimental range. Therefore, the supersymmetric CP violating
phases may affect the extraction of the angle $\gamma$. As an
example, let us consider the case where $R_B$ is enhanced from
$0.05$ (SM value) to $0.1$ and the phase $\phi_B$ is given by
$70^{\circ}$, which can be obtained by $\gamma \sim \pi/3$ and
$\phi_1\sim \pi/2$ or $\gamma \sim \pi/2$ and $\phi_1\sim \pi/3$.
In this case, one finds that \be R_+ \simeq  1.1, ~~~~ ~~~~~~ R_-
\simeq 0.94 ,~~~~ ~~~~~~ A_+ \simeq - A_{-} \simeq 0.2. \ee
Therefore, we can conclude that the SUSY contributions to $B^- \to
D K^-$ imply that $A_+$ and $A_-$ are within their $1\sigma$
experimental range simultaneously, unlike the SM results.

Finally, it is important to mention that in this scenario it is a
challenge to find a realistic SUSY model that accommodates these
results and satisfies all other constraints. Also the observation
of $A_+$ indicates that the ratio of the amplitudes for the
processes $B^- \to \bar{D}^0 K^-$ and $B^- \to D^0 K^-$ is larger
than $0.1$ which is rather difficult to obtain in the SM, so it
may be a hint for a new physics effect.
%
\subsection{{\large \bf $R_{\pm}$ and $A_{\pm}$ with large $D^0 -\bar{D}^0$ mixing}}

In the previous analysis, we have ignored the effect of the
$D-\bar{D}^0$ mixing. Now we consider this effect and define the
time dependent meson state $\vert D_1 \rangle \equiv \vert
D^0(t)\rangle$ and $\vert D_2 \rangle \equiv \vert
\bar{D}^0(t)\rangle$ as
 \bea
\vert D_1 \rangle &=& g_+(t) \vert D^0 \rangle +
\frac{q}{p}~g_-(t) \vert \bar{D}^0 \rangle~,\\
\vert D_2\rangle &=& g_+(t) \vert \bar{D}^0 \rangle +
\frac{p}{q}~g_-(t) \vert D^0 \rangle~, \eea where $q/p$ is
defined, as in the previous section, by \be \frac{q}{p}=
\sqrt{\frac{{\mathcal M}_{12}^*}{{\mathcal M}_{12}}} =
e^{-2i\theta_c}. \ee As shown in Eq.(\ref{thetac}), the phase
$\theta_c$ is of order one . The functions $g_{\pm}(t)$ is given
by \cite{Silva:1999bd} \be g_{\pm} = \frac{1}{2} \left(e^{-\mu_1
t} \pm e^{-i \mu_2 t}\right), \ee with $\mu_{i} = M_{D_i} - i
\Gamma_{D_i}/2$. In terms of $x_D= \frac{\Delta M_D}{\Gamma}$ and
$y_D = \frac{\Delta\Gamma}{2\Gamma}$, where $\Gamma=\Gamma_{D_1}
+\Gamma_2$, one finds \bea g_+(t)&=& e^{(-i M_D t -\tau/2)}
\left[1+(x_D -i y_D)^2 \tau^2/4 +...\right],\\
g_-(t)&=& e^{(-i M_D t -\tau/2)} \left[(-i x_D -y_D)^2 \tau/2
+...\right]. \eea Here $\tau = \Gamma t$. In this case, the decay
amplitudes of $B^- \to D K^-$ are given by \be A(B^- \to D_1 K^-)
= A(B^-\to D^0 K^-) g_+(t) + A(B^- \to \bar{D}^0 K^-)~
\frac{q}{p}~ g_-(t), \ee and \be A(B^- \to D_2 K^-) = A(B^-\to
\bar{D}^0 K^-) g_+(t) + A(B^- \to D^0 K^-)~ \frac{p}{q}~ g_-(t).
\ee Also the decay rates are defined as \cite{Silva:1999bd} \be
\Gamma(B^- \to D K^-) = \int dt~ \left\vert A(B^- \to D K^-)
\right\vert^2. \ee Therefore, one finds that \be \Gamma(B^- \to
D_1 K^-) = \vert A(B\to D^0 K^-)\vert^2 \left( G_+ + R_B^2 G_- + 2
R_B \mathrm{Re}\left[G_{+-} e^{-i(\delta_{B} +\phi_B - 2
\theta_c)} \right] \right), \ee where $G_i$ are given by \bea G_+
&=&\int_0^{\infty} \vert g_+(t) \vert^2 dt \simeq \frac{1}{\Gamma}
\left(1+\frac{y_D^2+x_D^2}{2}\right),\\
G_- &=&\int_0^{\infty} \vert g_-(t) \vert^2 dt \simeq
\frac{1}{\Gamma}
\left(\frac{y_D^2+x_D^2}{2}\right),\\
G_{+-} &=& \int_0^{\infty} g_+(t) g_{-}^*(t) dt \simeq
\frac{1}{\Gamma}\left(\frac{-y_D-i~x_D}{2}\right). \eea The CP
asymmetries $A_{CP_{1,2}}$ are defined by \be A_{CP_{1,2}}=
\frac{\Gamma(B^-\to D_{1,2} K^-) - \Gamma(B^+\to \bar{D}_{1,2}
K^+)}{\Gamma(B^-\to D_{1,2} K^-) + \Gamma(B^+\to \bar{D}_{1,2}
K^+)}. \ee Thus one can easily prove that \be A_{CP_{1}}=
\frac{R_B\left[y_D \sin \delta_B \sin(\phi_B -2 \theta_c) - x_D
\sin\delta_D \cos(\phi_B-2\theta_c)\right]}{G'_+ + R_B^2 G'_{-} -
R_B\left[y_D \cos\delta_B \cos(\phi_B - 2\theta_c) + x_D
\cos\delta_B \sin(\phi_B-2\theta_c)\right]}. \ee while \be
A_{CP_{2}}= \frac{R_B\left[y_D \sin \delta_B \sin(\phi_B -2
\theta_c) + x_D \sin\delta_D \cos(\phi_B-2\theta_c)\right]}{R_B^2
G'_+ + G'_{-} - R_B\left[y_D \cos\delta_B \cos(\phi_B - 2\theta_c)
- x_D \cos\delta_B \sin(\phi_B-2\theta_c)\right]}. \ee Here
$G'_{+,-}= \Gamma G_{+,-} =
(1+\frac{y_D^2-x_D^2}{2},\frac{y_D^2+x_D^2}{2})$ respectively. The
parameters $x_D$ and $y_D$ are subjected to stringent experimental
bounds in case of $\theta_c=0$,\cite{Nelson:1999bt}: $x_D^2 +y_D^2
\leq (6.7\times 10^{-2})^2$. For non-vanishing $\theta_c$, this
bound is no longer valid. However it is believed that in general
$x_D \sim y_D \sim 10^{-2}$. In this case, it is clear that $G'_+
\simeq 1$ and $G'_-\simeq 10^{-4}$ which imply that \be A_{CP_1}
\simeq 10^{-2} \times R_B \simeq {\cal O}(10^{-3}),\ee and \be
A_{CP_2} \simeq \frac{10^{-2}}{R_B} \simeq {\cal O}(0.1).\ee

From these results, it is remarkable that the effect of $D^0
-\bar{D}^0$ mixing breaks the usual relation between the CP
asymmetries $A_{CP_1}\equiv A_-$ and $A_{CP_2}\equiv A_+$: $A_+
\simeq - A_-$ which is satisfied in the SM and SUSY models with
negligible $D^0 -\bar{D}^0$ mixing, as we have emphasized in the
previous sections. As an example to show how natural to obtain in
this case CP asymmetries of order their central vales of the
experimental results in Eq.(\ref{Exp1}), let us consider $R_B
\simeq 0.15$, $x_D \simeq 3 \times 10^{-2}$, $y \simeq 5 \times
10^{-2}$, $\delta_B \sim \pi$ and $\phi_B \sim \theta_c\simeq
\pi/4$. One case easily find that \be A_{CP_1} \simeq
0.002,~~~~~~~~~~~~~ ~~~~~~ A_{CP_2} \simeq 0.3, \ee It is
interesting to note that these values of the CP asymmetries depend
on the CP violating SM phase $\gamma$ and the SUSY phase in the
$b\to u$ transition $\phi_1$, which contribute together to
$\phi_B$ as in Eq.(\ref{RB}), in addition to the $D^0-\bar{D}^0$
mixing phase $\theta_c$. Therefore, the determination of the angle
$\gamma$ relies on the new SUSY phases $\phi_1$ and $\theta_c$.
This confirms the fact the our determination of the SM angel might
be influenced by a new physics effect.

\section{{\large \bf Conclusions}}
In this paper we have studied supersymmetric contributions to $B^-
\to D^0 K^-$ and $B^- \to \bar{D}^0 K^-$ processes. We have shown
that in the SM, the branching ratios $R_{CP_{\pm}}$ of these
processes are within the experimental range. However the CP
asymmetry $A_{CP_+}$ is below its $1\sigma$ experimental lower
bound and the value of $A_{CP_-}$ is always negative. We have
performed a model independent analysis of the gluino and chargino
contributions to $b \to u$ and $b\to c$ transitions. We have used
the mass insertion approximation method to provide analytical
expressions for all the relevant Wilson coefficients.

The $D^0 -\bar{D}^0$ mixing experimental limits imply strong
constraints on the mass insertions $(\delta^u_{AB})_{12}$ which
affect the dominant gluino contribution to $B^- \to D K^-$. We
have revised these constraints and took them into account. We
showed that in case of negligible $D^0 - \bar{D}^0$ mixing, it is
possible to overcome these constraint and enhance the SUSY results
for the CP asymmetries in $B^- \to D K^-$ if one consider
simultaneous contributions from more than one mass insertion. In
this case, the $A_{CP_+}$ becomes within $1 \sigma$ experimental
range. However, with a large $D^0-\bar{D}^0$ mixing, one finds a
significant deviation between the two asymmetries and it becomes
natural to have them of order the central values of their
experimental measurements.

In general, We have emphasized that SUSY CP violating phases may
contribute significantly to the CP asymmetries in $B^- \to D K^-$
and therefore, they may affect our determination for the angle
$\gamma$ in the unitary triangle of the CKM mixing matrix.

%
\section*{{\large \bf Acknowledgements}}
I would like to thank G. Fasiel and E. Gabrielli for useful
discussions.


\end{document}